\documentclass[aps,twocolumn,floatfix,groupedaddress,nofootinbib]{revtex4}
\usepackage{csquotes}
\usepackage{graphicx,color}
\usepackage{float}
\usepackage[caption=false]{subfig}
\usepackage{amsmath}
\usepackage{amsmath}
\usepackage{multirow}
\usepackage{dsfont}
\usepackage[shortlabels]{enumitem}

\begin{document}

\title{Isomorph invariance and thermodynamics of repulsive dense bi-Yukawa one-component plasmas}
\author{F. Lucco Castello$^1$, P. Tolias$^1$, J. S. Hansen$^2$ and J. C. Dyre$^2$}
\affiliation{$^1$Space and Plasma Physics, Royal Institute of Technology, Stockholm, SE-100 44, Sweden\\
             $^2$Glass and Time, IMFUFA, Department of Science and Environment, Roskilde University, Roskilde, DK-4000, Denmark}
\begin{abstract}
\noindent In numerous realizations of complex plasmas, dust-dust interactions are characterized by two screening lengths and are thus better described by a combination of Yukawa potentials. The present work investigates the static correlations and the thermodynamics of repulsive dense bi-Yukawa fluids based on the fact that such strongly coupled systems exhibit isomorph invariance. The strong virial-potential energy correlations are demonstrated with the aid of molecular dynamics simulations, an accurate analytical expression for the isomorph family of curves is obtained and an empirical expression for the fluid-solid phase-coexistence line is proposed. The isomorph-based empirically modified hypernetted-chain approach, grounded on the ansatz of isomorph invariant bridge functions, is then extended to such systems and the resulting structural properties show an excellent agreement with the results of computer simulations. A simple and accurate closed-form expression is obtained for the excess internal energy of dense bi-Yukawa fluids by capitalizing on the compact parameterization offered by the Rosenfeld-Tarazona decomposition in combination with the Rosenfeld scaling, which opens up the energy route to thermodynamics.
\end{abstract}
\maketitle

\section{Introduction}\label{intro}

\noindent Complex (or dusty) plasmas can be defined as weakly ionized plasmas that are dispersed with solid charged particulates of mesoscopic size\,\cite{introd1,introd2}. For large dust densities and high charges, the system can be considered as one-component, since the dust species is dominant in terms of energy and momentum transfer\,\cite{introd1,introd2}. In the spatiotemporal scales relevant for dust dynamics, plasma assumes the role of the neutralizing background responsible for the quasi-stationary dust charge and the shielding of the dust-dust interaction. Such systems are typically engineered in the laboratory and, with proper manipulation, can virtually attain all states of matter\,\cite{introd2,introd3,introd4}. In the dense fluid region of the complex plasma phase diagram, the dust interaction energy is much higher than the dust kinetic energy but not high enough to drive crystallization.

In strongly coupled complex plasmas, regardless of dimensionality, the dust-dust interaction energy has been generally considered to conform to the Debye-H\"uckel or Yukawa type\,\cite{introd2,introd3,introd4,introd5}, $u(r)=(Q^2/r)\exp{(-r/\lambda_{\mathrm{D}})}$ where $Q$ is the dust charge and $\lambda_{\mathrm{D}}$ is the plasma screening length. This was supported by early measurements of the interaction between dust particles levitating in the sheath of a radio-frequency discharge\,\cite{introd6} and corresponds to the paradigm of the Yukawa one-component plasma (YOCP) which reduces to the classical one-component plasma (OCP) in the absence of screening ($\lambda_{\mathrm{D}}\to\infty$).

Yukawa interactions result from the linearized Boltzmann response of isotropic plasmas around a test point charge\,\cite{introd2}. However, even in the absence of plasma drifts, the continuous absorption of plasma fluxes on the particulate surface leads to non-Boltzmann densities for isolated dust\,\cite{potent0,potent1,potent2} and brings forth the issue of ionization-recombination balance for dense dust clouds\,\cite{potent3,potent4}. In the latter, simple hydrodynamic models have predicted that the interaction potential acquires a double Yukawa repulsive form\,\cite{potent5,potent6}, where the short-range screening length is controlled by plasma shielding and the long-range screening length is controlled by plasma sink-source competition. The possibility of long-range attraction has also been discussed in the literature\,\cite{potent7,potent8,potent9}. We shall refer to such strongly-interacting systems as bi-Yukawa one-component plasmas (biYOCP). It should be emphasized that the biYOCP interaction potential differs from the prototype two-Yukawa or double-Yukawa potential used to describe pair interactions in colloid-polymer mixtures, which also involves a hard-core repulsion\,\cite{gelati1,gelati2,gelati3}.

The general form of the bi-Yukawa pair-interaction potential is $u(r)=(\epsilon_1/r)\exp{(-r/\lambda_1)}+(\epsilon_2/r)\exp{(-r/\lambda_2)}$. We shall limit ourselves to purely repulsive interactions ($\epsilon_1,\epsilon_2>0$) with a pure Coulomb short-range asymptote ($\epsilon_1+\epsilon_2=Q^2$) and a dominant short-range screening term ($\lambda_2\geq\lambda_1,\epsilon_2\leq\epsilon_1$). We will focus exclusively on extended three-dimensional dense biYOCP liquids. Expressed in reduced coordinates, $x=r/d$ where $d=\left(4\pi{n}/3\right)^{-1/3}$ is the Wigner-Seitz radius with $n$ the particle density, the biYOCP interaction potential becomes
\begin{equation}
\beta{u}(x)=\frac{\Gamma}{x}\left[(1-\sigma)\exp{\left(-\kappa{x}\right)}+\sigma\exp{\left(-\mu\kappa{x}\right)}\right]\,.
\end{equation}
In the above $\beta=1/(k_{\mathrm{B}}T)$, where $T$ is the particle temperature and $k_{\mathrm{B}}$ is the Boltzmann constant. The dimensionless thermodynamic state variables of the biYOCP are the coupling parameter defined by $\Gamma=\beta{Q}^2/d$ and the normalized screening parameter defined by $\kappa=d/\lambda_1$. The external potential parameters of the biYOCP are the relative strength $\sigma=\epsilon_2/{Q}^2$ and the relative screening length $\mu=\lambda_1/\lambda_2$ subject to the constraints $0\leq\sigma\leq0.5$ and $0\leq\mu\leq1$. We shall assume that, in contrast to the state variables $(\Gamma,\kappa)$, the external parameters $(\sigma,\mu)$ depend on neither the density nor the temperature of the dust component. These parameters should vary in different experimental realizations, since they depend on the discharge conditions as well as on the dust size and composition. It is evident that the biYOCP system reduces to the YOCP for $\sigma=0$
\begin{equation}
\beta{u}(x)=\frac{\Gamma}{x}\exp{\left(-\kappa{x}\right)}\,,
\end{equation}
and that it reduces to the OCP for $\kappa=0$
\begin{equation}
\beta{u}(x)=\Gamma/x\,.
\end{equation}

In spite of the abundance of theoretical studies that predict a bi-Yukawa interaction potential for many laboratory realizations of complex plasmas\,\cite{potent5,potent6,potent7,potent8,potent9}, there have been few investigations of the thermodynamic, static and dynamic properties of the strongly coupled biYOCP\,\cite{biYuka1,biYuka2,biYuka3}. This probably stems from the fact that the two state variables and two external potential parameters form a four-dimensional $(\Gamma,\kappa;\sigma,\mu)$ space that is difficult to cover systematically. However, given the exhaustive study of the YOCP during the last decades as well as the anticipation that the dense biYOCP exhibits approximate hidden scale invariance and possibly obeys crucial additivity theorems, this number of independent parameters does not preclude the existence of simple closed-form relations for some fundamental aspects of dense biYOCP liquids.

The aim of the present paper is to provide simple analytical expressions that accurately describe the excess internal energy of dense repulsive bi-YOCP liquids for arbitrary $(\Gamma,\kappa;\sigma,\mu)$ parameters, which also opens up the energy route to all thermodynamic properties. The treatment is based on calculating the excess internal energy with the recently proposed isomorph-based empirically modified hypernetted-chain (IEMHNC) approximation\,\cite{approx3} and on taking advantage of the compact parameterization offered by the Rosenfeld-Tarazona decomposition\,\cite{RTdeco1} in combination with the Rosenfeld scaling\,\cite{DecRos1}.

The methodology should be highly accurate provided that dense repulsive biYOCP liquids are Roskilde simple systems, \emph{i.e.} that they possess continuous paths of constant excess entropy in their phase diagram along which a large number of properties in reduced units is invariant to a good approximation\,\cite{Roski01,Roski02,Roski03,Roski04,Roski05,Roski06,Roski07,Roski08,Roski09,Roski13}. Hence, the strong virial-potential energy correlations of the dense biYOCP are first demonstrated with the aid of molecular dynamics simulations, an analytical expression for the biYOCP isomorph family of curves is obtained and an empirical expression for the biYOCP melting line is introduced. These characteristics are also justified in light of their YOCP counterparts with the aid of approximate additivity theorems. The IEMHNC integral approach, based on the conjecture of isomorph invariant bridge functions, is then extended to biYOCP systems and the resulting structural properties are compared to molecular dynamics results. The Rosenfeld-Tarazona decomposition of the reduced excess internal energy\,\cite{RTdeco1} is finally generalized to dense biYOCP liquids; the static component is computed from the asymptotic high-density limit with hard-sphere Percus-Yevick radial distribution functions, while the thermal component is computed from the IEMHNC approach and fitted to the Rosenfeld scaling\,\cite{DecRos1}.

\section{Isomorph theory}\label{Isotheory}

\noindent Isomorphs are curves in the phase diagram along which a large set of the structural but also dynamic properties are approximately invariant when expressed in properly reduced units\,\cite{Roski01,Roski02,Roski03,Roski04,Roski05,Roski06,Roski07,Roski08,Roski09,Roski13}, in which the length is normalized to the mean-cubic inter-particle distance $a=n^{-1/3}$, the energy to $k_{\mathrm{B}}T$ and the time to $\tau=n^{-1/3}\sqrt{m/(k_{\mathrm{B}}T)}$, where $m$ is the particle mass. For one-component systems, isomorphs are of particular interest because they effectively transform  the phase diagram from two-dimensional to one-dimensional. Such curves exist for a rather sizable class of condensed matter systems, which are known as Roskilde-simple or R-simple.

\subsection{Definition of R-simple systems}\label{IsotheoryDef}

\noindent We first introduce the collective N-particle position vector configuration $\boldsymbol{R}=(\boldsymbol{r}_1,\boldsymbol{r}_2,...,\boldsymbol{r}_{\mathrm{N}})$, the total potential energy $U(\boldsymbol{R})$ and the microscopic virial $W(\boldsymbol{R})=(1/3)\sum_{i}\boldsymbol{r}_i\cdot\boldsymbol{F}_i=-(1/3)\boldsymbol{R}\cdot\nabla{U}(\boldsymbol{R})$. R-simple systems are systems that are characterized by strong correlations between their virial and their potential energy fluctuations in constant-volume $(NVT)$ equilibrium, which are quantified via the standard Pearson coefficient defined as $R_{WU}=\langle\Delta{W}\Delta{U}\rangle/\sqrt{\langle(\Delta{W})^2\rangle\langle(\Delta{U})^2\rangle}$ where $\langle...\rangle$ denotes canonical ensemble averaging and $\Delta{A}=A-\langle{A}\rangle$ statistical fluctuations around the thermodynamic mean. The strong $W-U$ correlations are empirically delimited by the practical condition $R_{WU}\gtrsim0.9$\,\cite{Roski01,Roski02,Roski03,Roski04}. Such a definition enables a straightforward identification of R-simple systems with the aid of $NVT$ Molecular Dynamics (MD) simulations\,\cite{Roski02,Roski03}, but allows neither the rigorous derivation of their fundamental properties nor the numerical determination of the isomorph curves. We shall refer to it as the \emph{practical definition} of R-simple systems.

For a more concrete characterization, the so-called \emph{axiomatic definition} of R-simple systems has been formulated\,\cite{Roski06,Roski07}, according to which an R-simple system is defined by the relation
\begin{equation}
U(\boldsymbol{R}_a)<U(\boldsymbol{R}_b) \Longleftrightarrow  U(\lambda\boldsymbol{R}_a)<U(\lambda\boldsymbol{R}_b),\label{eq-Rsimple}
\end{equation}
where the equilibrium configurations $\boldsymbol{R}_\mathrm{a},\,\boldsymbol{R}_\mathrm{b}$ correspond to the same density (not necessarily configurations of the same temperature) with $\lambda$ an arbitrary positive uniform scaling factor (though in practice of the order of unity as we will discuss in what follows). The validity of Eq.(\ref{eq-Rsimple}) for all possible equal density configurations implies perfect W-U correlations ($R_{WU}\equiv1$) which is strictly true only for Euler homogeneous pair-interactions such as those realized in inverse-power law (IPL) systems. Consequently, for typical R-simple systems, Eq.(\ref{eq-Rsimple}) should always be understood to be valid in a more restricted sense, \emph{i.e.} for most physically relevant configurations.

It is evident that both definitions depend on the thermodynamic state and should be satisfied in an extended region of the system's phase diagram. Overall, R-simple systems are usually identified according to the practical definition and it is accepted that all properties derived within the axiomatic definition will be followed in some approximate fashion. Three basic properties of R-simple systems are the following\,\cite{Roski06,Roski07,Roski08,Roski09}: (1) Single phases possess isomorphs and the isomorph curves constitute configurational adiabats, lines of constant excess entropy, of the respective phase region. (2) Phase coexistence curves are themselves isomorphs to the first-order; small deviations from this have recently been predicted using isomorph theory\,\cite{Roski09}. (3) Two arbitrary configurations $\boldsymbol{R}_1,\boldsymbol{R}_2$ that are identical in reduced units belong to thermodynamic state points on the same isomorph and have approximately equal probabilities in the canonical ensemble, \emph{i.e.}
\begin{equation}
\exp\left[{-\beta_1U(\boldsymbol{R}_1)}\right]\cong C_{12}\exp\left[{-\beta_2U(\boldsymbol{R}_2)}\right].\label{eq-EqualProbabilities}
\end{equation}
where the $C_{12}$ factor depends on the two thermodynamic state points but not on the microscopic configurations.

\subsection{Direct isomorph check}\label{IsotheoryMD}

\noindent Different methods have been developed for the tracing of the isomorph curves of R-simple systems via MD simulations\,\cite{Roski02,Roski05,Roski08}. The so-called direct isomorph check is based on the equal canonical probabilities and works as follows\,\cite{Roski02}. Eq.\eqref{eq-EqualProbabilities} can be rewritten as
\begin{equation}
U(\boldsymbol{R}_2)\cong\frac{T_2}{T_1}U(\boldsymbol{R}_1)+D_{12}\,.\label{eq-DirectCheck}
\end{equation}
Since these two configurations should be identical in reduced units, we also have that $n_1^{1/3}\boldsymbol{R}_1=n_2^{1/3}\boldsymbol{R}_2$. Let us suppose a $(n_1,T_1)$ reference state and an isomorphic state $(n_2,T_2)$ of re-scaled density $n_2=(1/\lambda){n}_1$. The goal is to determine $T_2$. The potential energy $U(\boldsymbol{R}_1)$ is first extracted by an equilibrium $(n_1,T_1)$ simulation, the configuration is then re-scaled to $\boldsymbol{R}_2=\lambda\boldsymbol{R}_1$ and the potential energy $U(\boldsymbol{R}_2)$ is obtained. This procedure is repeated for several configurations and a $U(\boldsymbol{R}_1),\,U(\boldsymbol{R}_2)$ scatter plot is obtained. As a consequence of Eq.\eqref{eq-DirectCheck}, the points should be well-approximated by a straight line whose slope $\theta$ can be obtained with linear regression. This allows for the determination of $T_2$ from $T_2=\theta T_1$. It is evident that an isomorph line consisting of $M$ points can be simply traced by applying the above procedure $M$ times, always starting from the newly obtained isomorphic point.

Theoretically, the direct isomorph check should work for arbitrary density jumps, \emph{i.e.} for any positive scaling factor $\lambda$. However, the approximate nature of isomorph invariance for non-IPL systems and thus of Eq.\eqref{eq-DirectCheck} should be taken into account. This imposes a $\lambda\lesssim1.1$ restriction, which indicates that tracing is usually performed with density increments $(n_1-n_2)/n_1\lesssim10\%$\,\cite{Roski08}. We note that, for some interaction potentials, it is more practical to carry out MD simulations directly in reduced units by setting the temperature and density equal to unity. The direct isomorph check is then performed by controlling the length and energy parameters of the potential instead of the density and temperature\,\cite{Roski08}.

\section{Isomorphic aspects of the YOCP structure and thermodynamics}\label{YOCP_Rsimple}

\noindent We recapitulate basic isomorphic aspects concerning key structural and thermodynamic YOCP properties, which will later be generalized to biYOCP systems. The strong $W-U$ equilibrium correlations of the YOCP are documented, the analytical expression of the YOCP isomorph family of curves is presented, the isomorph-based empirically modified hypernetted-chain approximation is introduced and the Rosenfeld-Tarazona decomposition of the internal energy is discussed.

\subsection{Strong correlations and isomorph curves}\label{YOCP_Rsimple_Corr}

\noindent A recent dedicated MD investigation has revealed that Yukawa liquids are R-simple systems providing a general framework for the understanding of earlier discovered invariant YOCP aspects\,\cite{Roski08}. The YOCP exhibits a high correlation coefficient $R_{WU}>0.98$ for an extended region of its liquid phase and YOCP isomorphs are accurately described by an analytical expression of the form\,\cite{Roski08}
\begin{equation}
\Gamma_{\mathrm{iso}}^{\mathrm{YOCP}}(\Gamma,\kappa)=\Gamma{e}^{-\alpha\kappa}\left[1+\alpha\kappa+\frac{(\alpha\kappa)^2}{2}\right]=\mathrm{const.}\,,\label{eq-YOCPmapping}
\end{equation}
where $\alpha=a/d=(4\pi/3)^{1/3}$ is the ratio of the cubic mean inter-particle distance to the Wigner-Seitz radius.

This expression is equivalent to the semi-empirical relation earlier proposed for the YOCP melting line\,\cite{isoKhr1,isoKhr2} that well describes the extensive MD liquid-solid phase coexistence data available in the literature\,\cite{Hamagu1,Hamagu2}
\begin{equation}
\Gamma_\mathrm{m}^{\mathrm{YOCP}}(\kappa)=\Gamma_{\mathrm{m}}^{\mathrm{OCP}}e^{\alpha\kappa}\left[1+\alpha\kappa+\frac{(\alpha\kappa)^2}{2}\right]^{-1}\,.\label{eq-YOCPmelting}
\end{equation}
In the above $\Gamma_{\mathrm{m}}^{\mathrm{OCP}}=171.8$ is the coupling parameter at the OCP melting point\,\cite{Hamagu1}.

\subsection{Isomorph-based integral theory approach}\label{YOCP_Rsimple_IEMHNC}

\noindent For isotropic one-component systems, the integral equation theory of liquid structure comprises of the Ornstein-Zernike equation\,\cite{bookre1}
\begin{equation}
h(r)=c(r)+n\int\,c(r^{\prime})h(|\boldsymbol{r}-\boldsymbol{r}^{\prime}|)d^3r^{\prime}\,,\label{OZequation}
\end{equation}
combined with the exact closure equation\,\cite{bookre1}
\begin{equation}
g(r)=\exp{\left[-\beta{u}(r)+h(r)-c(r)+B(r)\right]}\,,\label{OZclosure}
\end{equation}
where $h(r)=g(r)-1$ is the total correlation function, $g(r)$ the radial distribution function, $c(r)$ the direct correlation function and $B(r)$ the unknown bridge function. Different theoretical approaches employ different approximations for the bridge function, for instance $B(r)=0$ corresponds to the hypernetted-chain (HNC) approximation and $B(r)=\ln[1+\gamma(r)]-\gamma(r)$ corresponds to the Percus-Yevick (PY) approximation\,\cite{bookre1}, where we introduced the indirect correlation function $\gamma(r)=h(r)-c(r)$ for convenience.

The isomorph-based empirically modified hypernetted-chain (IEMHNC) approximation is based on the conjecture that bridge functions remain completely invariant along isomorph curves\,\cite{approx3}. For strongly coupled Yukawa systems, this recently developed approach assumes that the YOCP bridge function $B_{\mathrm{YOCP}}(r,\Gamma,\kappa)$ is an isomorph invariant up to and including the OCP limit and takes advantage of the availability of simulation-extracted OCP bridge functions $B_{\mathrm{OCP}}(r,\Gamma)$. From Eq.\eqref{eq-YOCPmapping} we have
\begin{equation}
B_{\mathrm{YOCP}}\left(\frac{r}{d},\Gamma,\kappa\right)=B_{\mathrm{OCP}}\left(\frac{r}{d},\frac{\Gamma}{e^{\alpha\kappa}}\left[1+\alpha\kappa+\frac{(\alpha\kappa)^2}{2}\right]\right)\,.\label{eq-bridge7}
\end{equation}
Accurate closed-form expressions are available for the OCP bridge function $B_{\mathrm{OCP}}(r,\Gamma)$ that have been obtained with the aid of Monte Carlo simulations and read as\,\cite{OCPbri1}
\begin{align}
&B_{\mathrm{OCP}}(r,\Gamma)=\Gamma\left[-b_0(\Gamma)+c_1(\Gamma)\left(\frac{r}{d}\right)^4+c_2(\Gamma)\left(\frac{r}{d}\right)^6\right.\nonumber\\&\,\qquad\qquad\quad\,\,\,\,\left.+c_3(\Gamma)\left(\frac{r}{d}\right)^8\right]\exp{\left[-\frac{b_1(\Gamma)}{b_0(\Gamma)}\left(\frac{r}{d}\right)^2\right]}\,,\label{eq-bridge1}
\end{align}
where the coefficients $b_0(\Gamma)$, $b_1(\Gamma)$, $c_1(\Gamma)$, $c_2(\Gamma)$ and $c_3(\Gamma)$ are described by Eqs.(21,23) of Ref.\cite{OCPbri1}. The IEMHNC approach has been validated against extensive YOCP computer simulations\,\cite{approx3} performing consistently better than other rigorous integral theory approaches\,\cite{approx4,approx6,approx7,approx8} as far as thermodynamic properties and structural quantities are concerned. Overall, it has been concluded that within its range of validity, $5.25\leq\Gamma_{\mathrm{iso}}^{\mathrm{YOCP}}(\Gamma,\kappa)\leq171.8$, the IEMHNC radial distribution functions are accurate within $1.2\%$ inside the first co-ordination cell as well as that the IEMHNC excess internal energies and pressures are accurate within $0.5\%$\,\cite{approx3}.

\subsection{The Rosenfeld-Tarazona decomposition of the internal energy}\label{YOCP_Rsimple_Rosenfeld}

\noindent On the basis of density functional theory and thermodynamic perturbation theory around the hard sphere reference system, by retaining the leading order terms of an asymptotic high-density expansion, Rosenfeld and Tarazona obtained a simple general decomposition of the excess internal energy of fluids valid at high densities and for predominant repulsive interactions\,\cite{RTdeco1}. In standard $(n,T)$ state variables, the RT decomposition reads as
\begin{equation}
U_{\mathrm{ex}}(n,T)=A(n)+B(n)T^{3/5}\,,\label{Roskildecomp}
\end{equation}
which leads to the well-known $T^{-2/5}$ RT scaling for the excess isochoric heat capacity $C_{\mathrm{V}}^{\mathrm{ex}}=\left(\partial{U}_{\mathrm{ex}}/\partial{T}\right)_{V}$. The RT scaling has been demonstrated to be closely followed by R-simple liquids\,\cite{RTdeco2}. It is worth pointing out that the original formulation of isomorph theory proved that $C_{\mathrm{V}}^{\mathrm{ex}}$ is isomorph invariant\,\cite{Roski02}, but this does not apply in the more generic reformulation\,\cite{Roski06} and has been confirmed not to apply for all R-simple liquids\,\cite{RTdeco3,RTdeco4}.

For the YOCP system, when expressed in $(\Gamma,\kappa)$ state variables and via reduced excess internal energies ${u}_{\mathrm{ex}}=U_{\mathrm{ex}}/(Nk_{\mathrm{B}}T)$, the RT decomposition reads as\,\cite{RTdeco1}
\begin{equation}
u_{\mathrm{ex}}(\Gamma,\kappa)=u_{\mathrm{st}}(\Gamma,\kappa)+u_{\mathrm{th}}(\Gamma,\kappa)\,,
\end{equation}
where $u_{\mathrm{ex}}(\Gamma,\kappa)=(3/2)\int_0^{\infty}x^2[\beta{u}(x;\Gamma,\kappa)]g(x;\Gamma,\kappa)dx$ in Wigner-Seitz coordinates $x=r/d$.

The \emph{static component} $u_{\mathrm{st}}$ refers to the excess internal energy of the system with its constituents frozen in a regular structure associated with the infinite-coupling limit. In this un-physical limit, the static correlations for repulsive pair additive systems coincide with the unitary packing fraction limit of the PY approximation for hard spheres\,\cite{RTdeco1}. For the YOCP we have
\begin{equation}
u_{\mathrm{st}}^{\mathrm{YOCP}}(\Gamma,\kappa)=6\Gamma\lim\limits_{\eta\rightarrow 1}\int_{0}^{\infty}yg_{\mathrm{PY}}(y;\eta)e^{-2\eta^{1/3}\kappa y}dy\,,\label{eq-ust}
\end{equation}
with $\eta=(\pi/6)nr_{\mathrm{hs}}^3$ the packing fraction, $r_{\mathrm{hs}}$ the hard sphere radius and $y=r/r_{\mathrm{hs}}$. The integral involving the PY hard-sphere radial distribution function $g_{\mathrm{PY}}(y;\eta)$ has an analytical form\,\cite{RTdeco5,RTdeco6} that, in the limit $\eta\rightarrow1$, yields
\begin{equation}
u_{\mathrm{st}}^{\mathrm{YOCP}}(\Gamma,\kappa)=\frac{\kappa(\kappa+1)\Gamma}{(\kappa+1)+(\kappa-1)e^{2\kappa}}\,,\label{eq-ustYOCP}
\end{equation}
which has a removable divergence in the OCP $\kappa\to0$ limit $u_{\mathrm{st}}^{\mathrm{YOCP}}(\Gamma,\kappa)=(3/2)\Gamma/\kappa^2-9\Gamma/10+\mathcal{O}(\kappa)$ that cancels out exactly with the background contribution leading to the well-known OCP result $u_{\mathrm{st}}^{\mathrm{OCP}}=-9\Gamma/10$\,\cite{RTdeco7,RTdeco8}. Note that Eq.\eqref{eq-ustYOCP} can also be derived within the Ion-Sphere Model (ISM)\,\cite{RTdeco7,RTdeco8} extended to the YOCP\,\cite{RTdeco9,RTdeco0}. The ISM approach is based on purely electrostatic arguments and offers a physically transparent proof but does not allow for straightforward generalizations to more complicated interaction potentials.

The \emph{thermal component} $u_{\mathrm{th}}$ accounts for deviations from fixed positions as caused by thermal motion. It can be acquired from Eq.(\ref{Roskildecomp}), that for IPL and Yukawa potentials, we have $u_{\mathrm{th}}\propto\Gamma^{2/5}$\,\cite{RTdeco1}, while a detailed analysis of YOCP MD simulations by Rosenfeld concluded that, when $\Gamma/\Gamma_\mathrm{m}^{\mathrm{YOCP}}\geq0.1$, $u_{\mathrm{th}}$ is well approximated by\,\cite{DecRos1}
\begin{equation}
u_{\mathrm{th}}^{\mathrm{YOCP}}(\Gamma,\kappa)=3.0\left[\frac{\Gamma}{\Gamma_{\mathrm{m}}^{\mathrm{YOCP}}(\kappa)}\right]^{2/5}\,.\label{eq-uthYOCP}
\end{equation}
We point out that the $\kappa-$dependence of the thermal component stems exclusively from the melting line $\Gamma_{\mathrm{m}}^{\mathrm{YOCP}}(\kappa)$. In view of Eqs.(\ref{eq-YOCPmapping},\ref{eq-YOCPmelting}), the above expression suggests that, for $\Gamma/\Gamma_\mathrm{m}^{\mathrm{YOCP}}\geq0.1$, the YOCP belongs to a subclass of R-simple liquids characterized by an isomorph invariant $C_{\mathrm{V}}^{\mathrm{ex}}$. Finally, we note that an empirical expression similar to Eq.\eqref{eq-uthYOCP}, reading as $u_{\mathrm{th}}^{\mathrm{YOCP}}=3.2({\Gamma}/{\Gamma_{\mathrm{m}}^{\mathrm{YOCP}}})^{2/5}-0.1$, gives a highly accurate prediction for the thermal component down to $\Gamma/\Gamma_\mathrm{m}^{\mathrm{YOCP}}=10^{-3}$\,\cite{RTdeco0}.

\section{Isomorphic aspects of the biYOCP structure and thermodynamics}\label{biYOCP_Rsimple}

\subsection{Strong correlations and isomorph curves}\label{biYOCP_Rsimple_Corr}

\noindent It can be argued that if $u_1(r),u_2(r)$ are the pair potentials of two R-simple liquids, then the pair potential $u_1(r)+u_2(r)$ should also correspond to a R-simple liquid\,\cite{Roski04,Roski10}. This additivity theorem, based on the quasi-universality of simple liquids\,\cite{Roski07}, clearly suggests that the dense biYOCP system is R-simple. Furthermore, Eq.(\ref{eq-EqualProbabilities}) suggests that the isomorph curves are also additive when expressed in $(\Gamma,\kappa)$ state variables. Therefore, the biYOCP isomorph curves will be given by $\Gamma_{\mathrm{iso}}^{\mathrm{biYOCP}}(\Gamma,\kappa;\sigma,\mu)= \Gamma_{\mathrm{iso}}^{\mathrm{YOCP}}[\Gamma(1-\sigma),\kappa]+\Gamma_{\mathrm{iso}}^{\mathrm{YOCP}}(\Gamma\sigma,\mu\kappa)$. Courtesy of Eq.\eqref{eq-YOCPmapping}, this leads to the analytical expression
\begin{align}
\Gamma_{\mathrm{iso}}^{\mathrm{biYOCP}}&(\Gamma,\kappa;\sigma,\mu)=\Gamma\left\{(1-\sigma){e}^{-\alpha\kappa}\left[1+\alpha\kappa+\frac{(\alpha\kappa)^2}{2}\right]\nonumber\right.\\&\left.+\sigma{e}^{-\mu\alpha\kappa}\left[1+\mu\alpha\kappa+\frac{(\mu\alpha\kappa)^2}{2}\right]\right\}=\mathrm{const}\label{eq-biYOCPmapping}
\end{align}
Finally, invoking Rosenfeld's additivity of melting temperatures\,\cite{Rosenf1,Rosenf2} (explained by isomorph theory\,\cite{Roski10}) and re-adjusting the coupling parameter strengths, we have $1/\Gamma_{\mathrm{m}}^{\mathrm{biYOCP}}(\kappa;\sigma,\mu)=(1-\sigma)/\Gamma_{\mathrm{m}}^{\mathrm{YOCP}}(\kappa)+\sigma/\Gamma_{\mathrm{m}}^{\mathrm{YOCP}}(\mu\kappa)$. Combining with Eq.(\ref{eq-YOCPmelting}), we get the biYOCP melting line
\begin{align}
\Gamma_{\mathrm{m}}^{\mathrm{biYOCP}}(\kappa;\sigma,\mu)&=\Gamma_{\mathrm{m}}^{\mathrm{OCP}}\bigg/\left\{\frac{(1-\sigma)}{{e}^{\alpha\kappa}}\left[1+\alpha\kappa+\frac{(\alpha\kappa)^2}{2}\right]\nonumber\right.\\&\left.+\frac{\sigma}{{e}^{\mu\alpha\kappa}}\left[1+\mu\alpha\kappa+\frac{(\mu\alpha\kappa)^2}{2}\right]\right\}.\label{eq-biYOCPmelting}
\end{align}

\begin{figure}[!b]
\centering
\includegraphics[width=3.3in]{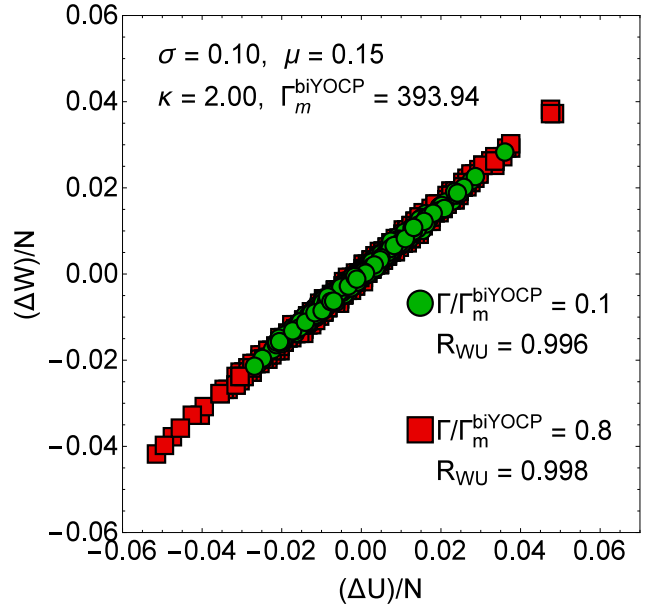}
\caption{(Color online) Scatter plot demonstrating the strong $W-U$ correlations of the biYOCP system for $\sigma=0.10,\,\mu=0.15$ at the state points $\kappa=2,\Gamma=0.1\Gamma_{\mathrm{m}}$ \& $\kappa=2,\Gamma=0.8\Gamma_{\mathrm{m}}$. The virial and potential energy equilibrium fluctuations have been normalized by the particle number $N$; $\kappa=2,\Gamma=0.1\Gamma_{\mathrm{m}}$ is characterized by $R_{WU}=0.996$ (green circles) \& $\kappa=2,\Gamma=0.8\Gamma_{\mathrm{m}}$ by $R_{WU}=0.998$ (red squares) revealing that the biYOCP is R-simple at both thermodynamic states.}\label{fig-strongcorrelations}
\end{figure}
As emphasized in Sec.\ref{IsotheoryDef}, the properties of R-simple systems are satisfied in an approximate manner for non-IPL potentials. Hence, MD simulations should be carried out in order to verify that biYOCP systems are indeed R-simple and to quantify the accuracy of the analytical expression for the biYOCP isomorphs. The MD simulations were carried out on graphics cards with the RUMD open-source software\,\cite{Roski11}. The simulations were performed directly in reduced units with $N=8192$ particles and a time-step of $\Delta{t}=2.5\times10^{-3}\tau$. The interaction potential was truncated at the cutoff distance $r_{\mathrm{cut}}=10a$ with the shifted-force cutoff method\,\cite{Roski12}. Since the Ewald decomposition is not implemented in RUMD, long-range interactions close to the unscreened Coulomb limit cannot be handled accurately\,\cite{Ewald00,Ewald01} and we confined ourselves to $\kappa\geq1$. We also focused on the dense fluid region of the phase diagram, $0.1<\Gamma/\Gamma_{\mathrm{m}}^{\mathrm{biYOCP}}<1$.

The biYOCP system exhibited a very high correlation coefficient for all state points studied. The strong correlations between the virial and potential energy fluctuations are documented in figure \ref{fig-strongcorrelations} for $\sigma=0.10,\,\mu=0.15$ and two thermodynamic states, both characterized by $R_{WU}>0.99$. Overall, it is concluded that biYOCP liquids are R-simple for any $\sigma<0.5$, $\mu<1$ at least in the phase region defined by $0.1<\Gamma/\Gamma_{\mathrm{m}}^{\mathrm{biYOCP}}<1$ and $\kappa\geq1$.
\begin{figure}
\centering
\includegraphics[width=3.3in]{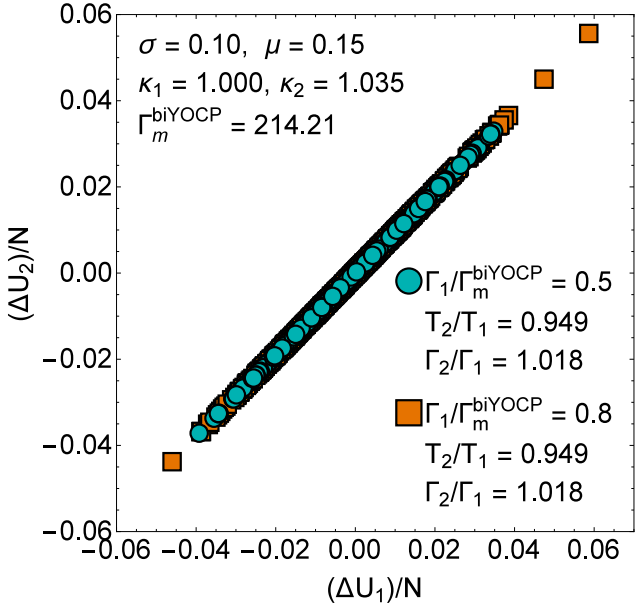}
\caption{(Color online) Direct isomorph check for the biYOCP system when $\sigma=0.10,\,\mu=0.15$. Scatter plot of the potential energy of two initial configurations vs the potential energy of two respective scaled configurations. The two initial configurations correspond to the state points $\kappa_1=1.0$, $\Gamma_1=0.5\Gamma_{\mathrm{m}}$ (cyan circles) \& $\kappa_1=1.0$, $\Gamma_1=0.8\Gamma_{\mathrm{m}}$ (orange squares), for both cases the scaled screening parameter is $\kappa_2=1.035$. The temperature $T_2$ of the scaled configurations is evaluated by multiplying the linear regression slope $\theta$ with the temperature $T_1$ of the initial configurations, \emph{i.e.} $T_2=\theta{T}_1$. The coupling parameter is then obtained from $\Gamma_2=(\kappa_1/\kappa_2)(\Gamma_1/\theta)$.}\label{fig-directiso}
\end{figure}

\begin{figure}
	\centering
	\includegraphics[width=3.4in]{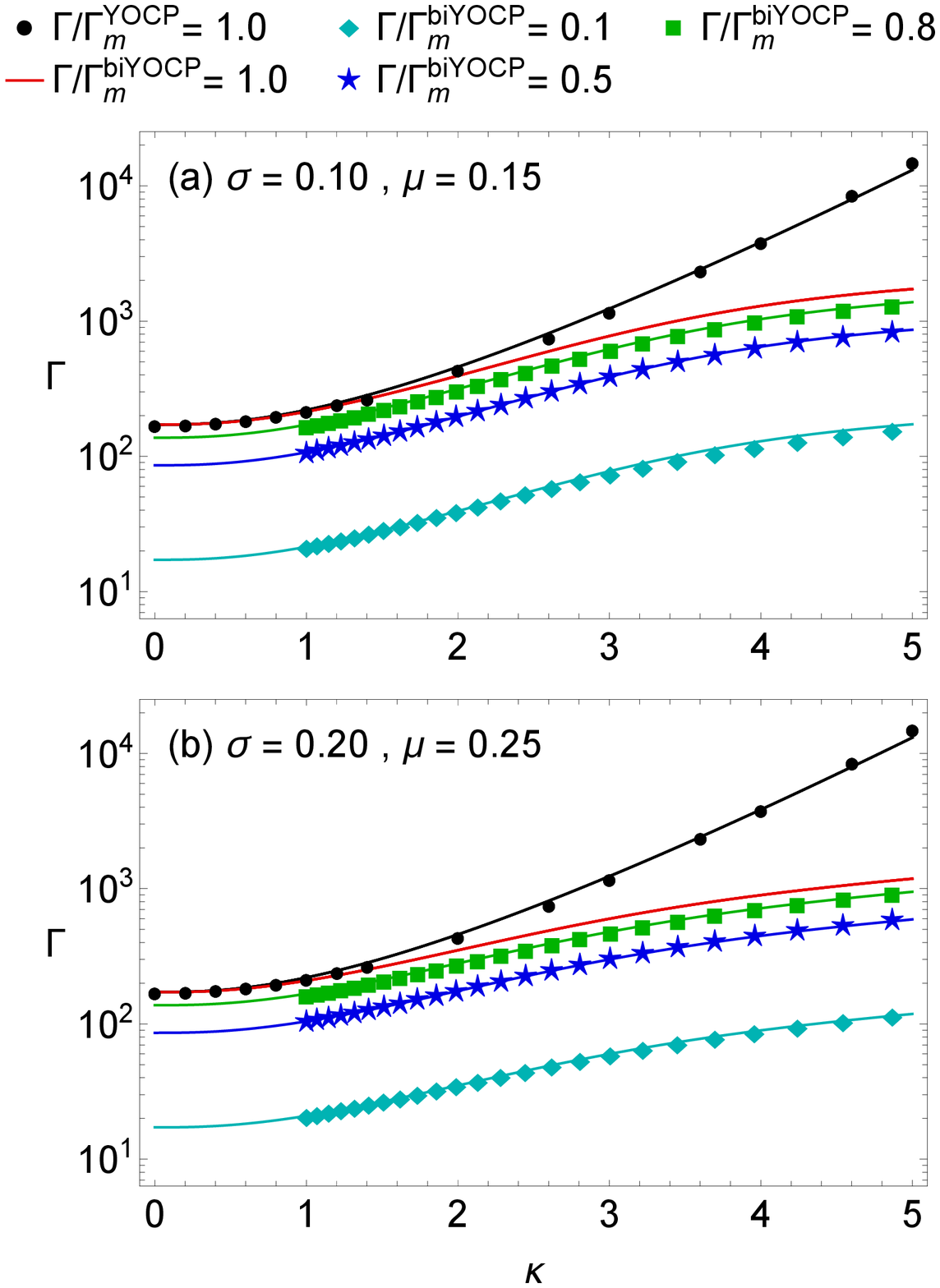}
	\caption{(Color online) Three isomorph curves together with the melting line for the dense biYOCP liquid in the $(\log{\Gamma},\kappa)$ phase diagram. The three isomorphs as obtained from the MD simulations (cyan, blue and green discrete points) are compared to the analytical approximations obtained with Eq.\eqref{eq-biYOCPmapping} (cyan, blue and green solid lines). The numerical isomorphs consist of $48$ points. In the figure, they were down-sampled to $24$ points in order to improve visibility. Each subplot focuses on a single $\sigma,\,\mu$ combination; namely (a) $\sigma=0.10,\,\mu=0.15$ and (b) $\sigma=0.20,\,\mu=0.25$. The biYOCP melting curves that result from Eq.\eqref{eq-biYOCPmelting} (red solid line) are also reported and compared to the YOCP melting line with black dots for the MD results of Ref.\cite{Hamagu2} and black solid lines for the analytical representation of Eq.\eqref{eq-YOCPmelting}.}\label{fig-isomorphs}
\end{figure}

The direct isomorph check with a screening parameter increment $\Delta\kappa/\kappa=3.5\%$ was employed in order to trace isomorphs in the interval $1.0\leq\kappa\leq5.0$. The screening parameters of the state points that are generated by the direct isomorph check belong to a geometric progression with a common ratio of $1.035$. The initial step of the numerical procedure is illustrated in figure \ref{fig-directiso} and detailed in the respective caption, see also Sec.\ref{IsotheoryMD}. The numerical isomorph curves are illustrated in figure \ref{fig-isomorphs} together with the analytical expression of Eq.\eqref{eq-biYOCPmapping}. We considered three reduced coupling parameters $\Gamma/\Gamma_{\mathrm{m}}^{\mathrm{biYOCP}}=\left\{0.1,0.5,0.8\right\}$ and two $(\sigma,\,\mu)$ combinations: $\sigma=0.10$, $\mu=0.15$ in Figure \ref{fig-isomorphs}a and  $\sigma=0.20$, $\mu=0.25$ in Figure \ref{fig-isomorphs}b.

The closed-form description of the isomorph lines, see Eq.\eqref{eq-biYOCPmapping}, displays a remarkable accuracy when compared with the discrete numerical isomorph points. The mean absolute relative deviations between the analytical and numerical isomorphs never exceeded $3.3\%$, thus verifying the additivity theorems. In the same figure, the melting line for the biYOCP has been compared to the melting line for the YOCP as obtained from MD simulations\,\cite{Hamagu2} and from the analytical expression of Eq.\eqref{eq-YOCPmelting}. As expected from the longer range of the interactions, for the same screening parameter, the melting point of the biYOCP is always smaller than the melting point of the YOCP, with the obvious exception of $\kappa=0$ where both systems reduce to the OCP.

\subsection{Isomorph-based integral theory approach}\label{biYOCP-integral}

\noindent The results of the previous subsection allow us to extend the IEMHNC approach to biYOCP systems. Taking into account the fact that the biYOCP reduces to the OCP for $\kappa\to0$ and invoking the ansatz of isomorph-invariant bridge functions, we can employ the biYOCP isomorph mapping introduced in Eq.\eqref{eq-biYOCPmapping} in order to acquire a correspondence relation between the unknown biYOCP bridge function and the known OCP bridge function\,\cite{approx3},
\begin{align}
&B_{\mathrm{biYOCP}}\left(\frac{r}{d},\Gamma,\kappa;\sigma,\mu\right)=B_{\mathrm{OCP}}\left(\frac{r}{d},\Gamma\left\{\frac{(1-\sigma)}{{e}^{\alpha\kappa}}\left[1+\nonumber\right.\right.\right.\\&\quad\,\left.\alpha\kappa+\frac{(\alpha\kappa)^2}{2}\right]\left.\left.+\frac{\sigma}{{e}^{\mu\alpha\kappa}}\left[1+\mu\alpha\kappa+\frac{(\mu\alpha\kappa)^2}{2}\right]\right\}\right).\label{eq-bridgebiYOCP}
\end{align}
The range of validity of the simulation-extracted OCP bridge function, Eq.(\ref{eq-bridge1}), dictates the validity range of the constructed biYOCP bridge function, Eq.(\ref{eq-bridgebiYOCP}), which becomes $5.25\leq\Gamma_{\mathrm{iso}}^{\mathrm{biYOCP}}(\Gamma,\kappa;\sigma,\mu)\leq171.8$.

The biYOCP radial distribution functions obtained from the IEMHNC approach and MD simulations will be compared to quantify the accuracy of the integral theory. The numerical procedure followed for the solution of the system of Eqs.(\ref{OZequation},\ref{OZclosure},\ref{eq-bridge1},\ref{eq-bridgebiYOCP}) is based on Picard iterations in Fourier space combined with standard mixing algorithms and established long-range decomposition schemes (when necessary) in order to improve convergence\,\cite{approx3}. The upper range cut-off was selected to be $R_{\mathrm{max}}=20\,d$, the real space resolution was $\Delta{r}=10^{-3}\,d$ and the Fourier-space convergence criterion in terms of the indirect correlation function read as $|\gamma_{n}(k)-\gamma_{n-1}(k)|<10^{-5}\,\forall{k}$. The MD simulations were performed with the RUMD software as described in section \ref{biYOCP_Rsimple_Corr}.

\begin{figure}
	\centering
	\includegraphics[width=3.4in]{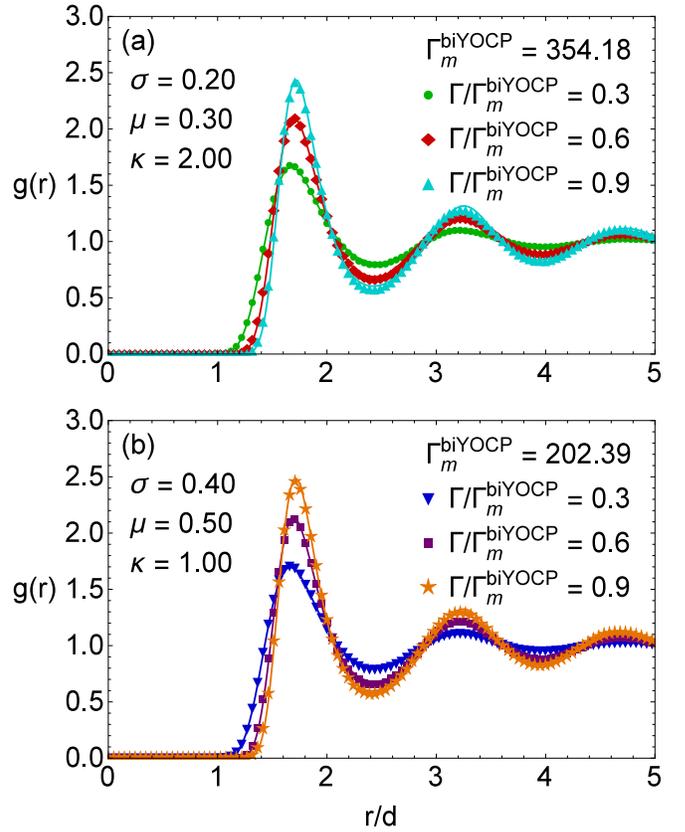}
	\caption{(Color online) Radial distribution functions obtained from MD simulations (discrete points) and from the IEMHNC approach (solid lines). (a) Results for $\sigma=0.20$, $\mu=0.30$, $\kappa=2.00$ and $\Gamma/\Gamma_{\mathrm{m}}^{\mathrm{biYOCP}}$=0.3 (green), $0.6$ (red), $0.9$ (cyan). The corresponding coupling parameter at the melting point is $\Gamma_{\mathrm{m}}^{\mathrm{biYOCP}}(\kappa;\sigma,\mu)=354.18$ according to Eq.(\ref{eq-biYOCPmelting}). (b) Results for $\sigma=0.40$, $\mu=0.50$, $\kappa=1.00$ and $\Gamma/\Gamma_{\mathrm{m}}^{\mathrm{biYOCP}}$=0.3 (blue), $0.6$ (purple), $0.9$ (orange). According to Eq.(\ref{eq-biYOCPmelting}), the coupling parameter at the melting point is $\Gamma_{\mathrm{m}}^{\mathrm{biYOCP}}(\kappa;\sigma,\mu)=202.39$. The agreement between theory and simulations is remarkable.}\label{fig-rdf}
\end{figure}

The IEMHNC and MD radial distribution functions for the biYOCP system have been compared in Figure \ref{fig-rdf} for six different combinations of external parameters $(\sigma,\mu)$ and phase variables $(\Gamma,\kappa)$. Despite the lack of adjustable parameters in the IEMHNC approach, the agreement between the theory and MD simulations is excellent with the resulting $g(r)$ being almost indistinguishable especially within the first coordination cell. For a more tangible comparison, the correlation void, \emph{i.e.} the distance where $g(r)=1/2$, and the magnitude of the first maximum were extracted. For all the state points investigated, the IEMHNC deviations from MD never exceeded $1\%$ for the correlation void, $2\%$ for the first maximum magnitude. The results indicate that the IEMHNC approach remains remarkably accurate also for biYOCP liquids and suggest that the IEMHNC structural and thermodynamic properties can be considered as nearly exact.

\subsection{The Rosenfeld-Tarazona decomposition of the internal energy}\label{biYOCP_Rsimple_Rosenfeld}

\noindent The Rosenfeld-Tarazona decomposition of the reduced excess internal energy for biYOCP liquids is naturally
\begin{align}
u_{\mathrm{ex}}^{\mathrm{biYOCP}}(\Gamma,\kappa;\sigma,\mu)=&u_{\mathrm{st}}^{\mathrm{biYOCP}}(\Gamma,\kappa;\sigma,\mu)+\nonumber\\&u_{\mathrm{th}}^{\mathrm{biYOCP}}(\Gamma,\kappa;\sigma,\mu)\,.\label{eq-ubiYOCP}
\end{align}
We should derive closed-form expressions for the static component $u_{\mathrm{st}}^{\mathrm{biYOCP}}(\Gamma,\kappa;\sigma,\mu)$ and the thermal component $u_{\mathrm{th}}^{\mathrm{biYOCP}}(\Gamma,\kappa;\sigma,\mu)$. As per usual, for the static part we need to consider the unitary packing fraction limit of the PY approximation, whereas for the thermal part we will consider the result of the IEMHNC approximation.

We begin with the \emph{static component}. For the biYOCP interaction potential, the infinite-coupling limit can be expressed as
\begin{align}
&u_{\mathrm{st}}^{\mathrm{biYOCP}}=6\Gamma(\sigma-1)\lim\limits_{\eta\rightarrow1}\int_{0}^{\infty}yg_{\mathrm{PY}}(y;\eta)e^{-2\eta^{1/3}\kappa{y}}dy\nonumber\\&\qquad\qquad+6\sigma\Gamma\lim\limits_{\eta\rightarrow1}\int_{0}^{\infty}yg_{\mathrm{PY}}(y;\eta)e^{-2\eta^{1/3}\kappa\mu{y}}dy.\label{eq-ustbiYOCP_infinitecoupling}
\end{align}
Owing to the presence of the PY radial distribution function for hard spheres at $\eta=1$ packing, there is an inherent additivity in the static component with each integral representing an independent YOCP contribution. Therefore, the static part for the biYOCP is simply given by
\begin{align}
u_{\mathrm{st}}^{\mathrm{biYOCP}}&(\Gamma,\kappa;\sigma,\mu)=\Gamma(1-\sigma)\frac{\kappa(\kappa+1)}{(\kappa+1)+(\kappa-1)e^{2\kappa}}\nonumber\\&\quad\quad\quad\,\,\,\,+\Gamma\sigma\frac{\mu\kappa(\mu\kappa+1)}{(\mu\kappa+1)+(\mu\kappa-1)e^{2\mu\kappa}}\,.\label{eq-ustbiYOCP}
\end{align}
We continue with the \emph{thermal component}. On the basis of Rosenfeld's theoretical and numerical analysis for the YOCP\,\cite{DecRos1}, it is expected that, for $\Gamma/\Gamma_\mathrm{m}^{\mathrm{biYOCP}}\geq0.1$, the thermal part will be described by a relation of the form
\begin{equation}
u_{\mathrm{th}}^{\mathrm{biYOCP}}(\Gamma,\kappa;\sigma,\mu)=\delta(\sigma,\mu)\left[\frac{\Gamma}{\Gamma_{\mathrm{m}}^{\mathrm{biYOCP}}(\kappa;\sigma,\mu)}\right]^{\frac{2}{5}}\label{eq-uthbiYOCP}
\end{equation}
with $\delta(\sigma,\mu)$ a function of the external potential parameters. The IEMHNC approach will be utilized to quantify the accuracy of this relation and to determine the unknown function $\delta(\sigma,\mu)$.

A systematic parametric study was performed, where the IEMHNC approximation was numerically solved over the whole range of physically meaningful external potential parameters $(\sigma,\mu)$, within the dense liquid phase $0.1\Gamma_\mathrm{m}^{\mathrm{biYOCP}}(\kappa;\sigma,\mu)\leq\Gamma\leq\Gamma_\mathrm{m}^{\mathrm{biYOCP}}(\kappa;\sigma,\mu)$ and for typical normalized screening parameters $0\leq\kappa\leq4.0$. In particular, we probed five equally stepped $\sigma$ values ranging from $0.05$ to $0.45$ and eleven $\mu$ values equally stepped in the $0.05$ to $0.95$ range including the Coulomb limit $\mu=0$. For each of the $55$ $(\sigma,\mu)$ combinations, we studied eight equally stepped screening parameters ranging from $0.5$ to $4.0$ and ten equally stepped coupling parameters. In total, $4400$ $(\sigma,\mu,\Gamma,\kappa)$ combinations were investigated, which are summarized in Table \ref{tab-parametricstudy}.
\begin{table}[t!]
	\caption{Parametric study for the determination of $\delta(\sigma,\mu)$ from the IEMHNC approach. For each parameter; $a_{\mathrm{min}}$ denotes the lowest value studied, $a_{\mathrm{max}}$ the highest value probed, $N_a$ the distinct number of values investigated, $\Delta{a}$ the constant step between successive values. Note that the Yukawa plus Coulomb limit corresponding to $\mu=0$ was also studied.}
	\label{tab-parametricstudy}
	\centering
	\begin{tabular}{ccccc}
		\hline
		Parameter                    & \,\,$a_\mathrm{min}$\,\, & \,\,$a_\mathrm{max}$\,\, & \,\,$N_a$\,\, & \,\,$\Delta a$\,\, \\ \hline			
		$\sigma$                     & 0.05                     & 0.45                     & 5             & 0.1 \\
		$\mu$                        & 0.05                     & 0.95                     & 10            & 0.1 \\
		$\Gamma/\Gamma_{\mathrm{m}}$ & 0.1                      & 1.0                      & 10            & 0.1 \\
		$\kappa$                     & 0.5                      & 4.0                      & 8             & 0.5 \\ \hline
	\end{tabular}
\end{table}

For each case, the IEMHNC radial distribution function was numerically computed and the reduced excess internal energy was calculated from $u_{\mathrm{ex}}^{\mathrm{biYOCP}}(\Gamma,\kappa;\sigma,\mu)=(3/2)\Gamma\int_0^{\infty}x[\sigma{e}^{-\kappa{x}}+(1-\sigma){e}^{-\mu\kappa{x}}]g(x;\Gamma,\kappa;\sigma,\mu)dx$ where $x=r/d$. Afterwards, the static part was calculated from Eq.(\ref{eq-ustbiYOCP}) and the thermal part was obtained from the decomposition $u_{\mathrm{th}}^{\mathrm{biYOCP}}(\Gamma,\kappa;\sigma,\mu)=u_{\mathrm{ex}}^{\mathrm{biYOCP}}(\Gamma,\kappa;\sigma,\mu)-u_{\mathrm{st}}^{\mathrm{biYOCP}}(\Gamma,\kappa;\sigma,\mu)$. Then, a two-step least-square fitting procedure was followed; (1) For each fixed $(\sigma_i,\mu_i)$, the corresponding $\delta$ value was determined by least-square fitting the $u_{\mathrm{th}}^{\mathrm{biYOCP}}(\Gamma,\kappa;\sigma_i,\mu_i)$ values obtained for all $(\Gamma,\kappa)$ combinations to Eq.(\ref{eq-uthbiYOCP}). This was repeated for all $(\sigma,\mu)$ combinations and a $\delta(\sigma_i,\mu_i)$ dataset was obtained. (2) The function $\delta(\sigma,\mu)$ was eventually extracted with a two-dimensional least-squares fit to the $\delta(\sigma_i,\mu_i)$ dataset.

\begin{figure}[t!]
	\centering
	\includegraphics[width=3.0in]{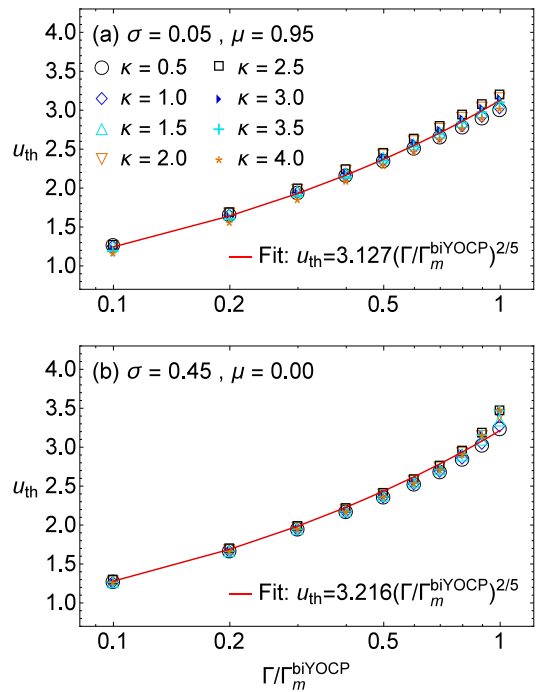}
	\caption{(Color online) The thermal component of the reduced excess internal energy $u_{\mathrm{th}}$ for dense biYOCP liquids as function of the reduced coupling parameter $\Gamma/\Gamma_{\mathrm{m}}^{\mathrm{biYOCP}}$ for two fixed $(\sigma,\mu)$ values; (a) $\sigma=0.05$, $\mu=0.95$, (b) $\sigma=0.45$, $\mu=0.00$. Eight normalized screening parameters $\kappa$ and ten coupling parameters $\Gamma$ have been probed for each $(\sigma,\mu)$. Discrete points illustrate the result of the IEMHNC approximation, whereas solid lines illustrate Rosenfeld's analytical expression, Eq.\eqref{eq-uthbiYOCP}, with a least-square fitted pre-factor $\delta$. In each subplot the $\delta$ value is reported in the lower right edge.}\label{fig-Gammakappafit}
\end{figure}

The results revealed that Rosenfeld's expression $u_{\mathrm{th}}\propto(\Gamma/\Gamma_{\mathrm{m}})^{2/5}$ provides an excellent description of the thermal internal energy component of dense biYOCP liquids regardless of the value of the external parameters $(\sigma,\mu)$. To be more quantitative, for any of the $55$ $(\sigma_i,\mu_i)$ combinations, the mean absolute relative deviations between the IEMHNC dataset (consisting of $80$ points) and the Eq.(\ref{eq-uthbiYOCP}) expression with a least-square fitted pre-factor $\delta(\sigma_i,\mu_i)$ never exceeded $2.0\,\%$. In fact, the highest deviations were recorded for $\mu=0.00$, \emph{i.e.} the Yukawa-plus-Coulomb special case, which is an expected result in view of the well documented observation that the $2/5$ exponent in Rosenfeld's expression becomes less accurate for Coulomb potentials\,\cite{RTdeco0,RTdeco8,fitKhra}. The collapse of the IEMHNC results to the analytical curve has been illustrated in figure \ref{fig-Gammakappafit} for two $(\sigma,\mu)$ combinations.
\begin{figure}[!ht]
	\centering
	\includegraphics[width=3.0in]{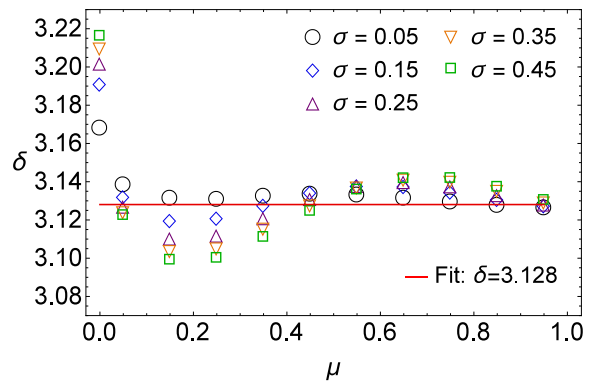}
	\caption{(Color online) The least-square fitted pre-factor $\delta$ of Rosenfeld's expression for the thermal component of the reduced excess internal energy $u_{\mathrm{th}}$ for dense biYOCP liquids, see Eq.(\ref{eq-uthbiYOCP}), as a function of the external parameters $\sigma<0.5$ and $\mu<1$. The $55$ discrete points cluster around a constant value, which is best represented by $\delta=3.128$ (red solid line).}\label{fig-deltafit}
\end{figure}

Concerning the $\delta(\sigma,\mu)$ function, no systematic dependencies on either $\sigma$ or $\mu$ were observed. As illustrated in figure \ref{fig-deltafit}, the least-square fitted $\delta$ pre-factor remained nearly constant as $\sigma,\mu$ vary. The largest deviations from the average value were attained for $\mu=0.00$. Excluding this case, we obtained $\delta(\sigma,\mu)=3.128$ for the best-fit solution with a merely $0.23\%$ mean relative error between the constant value and the discrete points $\delta(\sigma_i,\mu_i)$. This result is reminiscent of previous results obtained for the YOCP case ($\sigma=0$) which predicted $\delta$ to lie in the range $3.0\leq\delta\leq3.2$\,\cite{DecRos1,RTdeco0,fitKhrR}. Furthermore, the result is also accurate for $\mu=0.00$, for which it exhibits only a $0.75\%$ mean relative error with respect to the discrete points. Altogether, the thermal component of the reduced excess internal energy for dense biYOCP liquids is given by
\begin{equation}
u_{\mathrm{th}}^{\mathrm{biYOCP}}(\Gamma,\kappa;\sigma,\mu)=3.128\left[\frac{\Gamma}{\Gamma_{\mathrm{m}}^{\mathrm{biYOCP}}(\kappa;\sigma,\mu)}\right]^{\frac{2}{5}}\,.\label{eq-uthbiYOCP_fit}
\end{equation}
This general result is valid for arbitrary external parameters $\sigma<0.5$, $\mu<1$. It should be noted that the $\kappa,\sigma,\mu$ dependencies of the thermal component stem exclusively from the melting line $\Gamma_{\mathrm{m}}^{\mathrm{biYOCP}}(\kappa;\sigma,\mu)$. Excluding the Coulomb limit $\mu=0$, for the $4000$ $(\sigma,\mu\neq0,\Gamma,\kappa)$ combinations investigated, the mean relative deviation between Eq.(\ref{eq-uthbiYOCP_fit}) and the IEMHNC approach result is $1.4\%$ whereas the maximum relative deviation is $5.7\%$.

Combining all the above, the reduced excess internal energy for dense repulsive biYOCP liquids is described by Eqs.(\ref{eq-biYOCPmelting},\ref{eq-ubiYOCP},\ref{eq-ustbiYOCP},\ref{eq-uthbiYOCP_fit}). Excluding the Coulomb limit $\mu=0$, the mean relative deviation between Eqs.(\ref{eq-biYOCPmelting},\ref{eq-ubiYOCP},\ref{eq-ustbiYOCP},\ref{eq-uthbiYOCP_fit}) and the IEMHNC approximation result is $0.1\%$ whereas the maximum relative deviation is $3.3\%$ for the $4000$ $(\sigma,\mu\neq0,\Gamma,\kappa)$ combinations studied. Higher deviations are generally observed for large values of $\kappa$ and low values of $\Gamma/\Gamma_{\mathrm{m}}$, see also Ref.\cite{fitKhrR}. Given the four parameters involved in the biYOCP interaction, such an accurate but compact representation would have been impossible had we not taken advantage of known YOCP representations.

\section{Discussion and conclusions}

\noindent Repulsive bi-Yukawa pair-interactions have been theoretically predicted to adequately describe dust-dust interactions in complex plasmas that are engineered in weakly ionized gas discharges\,\cite{potent5,potent6,potent7,potent8,potent9}. In isotropic complex plasmas, the short-range characteristic length arises from the polarization of the plasma background and is close to the linear Debye-H\"uckel screening length, whereas the long-range characteristic length arises from the competition between plasma ionization and plasma absorption or recombination. Complex plasmas have served as the main motivation for the present investigation, but we should point out that there are only indirect evidence (obtained in fluid de-mixing experiments\,\cite{potentO}) which support such interaction model. Repulsive bi-Yukawa pair-interactions have also been utilized as effective ion-ion potentials for the regime of warm dense matter (WDM)\,\cite{WDMpap1,WDMpap2}, where strongly coupled classical ions are coexisting with degenerate electrons\,\cite{WDMpapg}. In WDM, the long-range characteristic length arises from the polarization of the free electron cloud and is very close to the linear Thomas-Fermi screening length, whereas the short-range characteristic length arises from the overlapping bound electron wavefunctions and leads to increased ion repulsion\,\cite{WDMpap1,WDMpap2,WDMpap3}. Such effective potentials have been extracted from density functional molecular dynamics simulations considering all electrons\,\cite{WDMpap3} and have also been indirectly confirmed by experiments\,\cite{WDMpap4,WDMpap5}.

Overall, the paradigm of bi-Yukawa one-component plasmas can be utilized for the study of the properties of strongly coupled systems whose inter-particle interactions are characterized by two fundamental length scales. Therefore, it is entirely plausible that such effective potentials emerge in a broader range of classical soft matter systems and degenerate subatomic systems than currently suspected. Theoretical investigations of dynamic, structural and thermodynamic properties of the dense biYOCP can greatly benefit from the comprehensive understanding of the strongly coupled YOCP. In particular, the approximate isomorph invariance of both model systems (together with the ensuing additivity theorems) provides a general framework that allows the straightforward extension of successful theoretical approaches and even of accurate semi-empirical expressions from the single to the double Yukawa potential.

Here, we focused on structural and mainly on thermodynamic aspects of dense repulsive biYOCP liquids. An accurate description for the isomorph family of phase-diagram curves has been obtained from molecular dynamics simulations, with the analytical expression constructed on the basis of approximate additivity theorems. The isomorph-based empirically modified hypernetted-chain approach, based on the isomorph invariance of the bridge functions and known to be highly accurate for the YOCP, has been extended to the biYOCP and the resulting structural properties have revealed an excellent agreement with the results of computer simulations. In spite of the involvement of two state variables and two external parameters in the pair-interaction potential, this allowed us to obtain a simple and accurate closed-form expression for the excess internal energy of dense biYOCP liquids by capitalizing on a compact parameterization procedure that was yet again initially established for the YOCP.

\section*{Acknowledgments}

\noindent FLC and PT acknowledge the financial support of the Swedish National Space Agency. This work was also supported by the VILLUM Foundation's Grant No.\,16515 (Matter).

\end{document}